\documentclass[letterpaper, 10 pt, conference]{ieeeconf}  

\IEEEoverridecommandlockouts                              

\usepackage{graphicx} 
\usepackage{subcaption} 
\usepackage{float} 
\usepackage{booktabs}
\usepackage{amsmath}
\usepackage{svg}
\usepackage{tabularx}
\usepackage{xcolor}
\usepackage{textcomp}
\usepackage{soul}
\usepackage{epsfig}
\usepackage{tikz}
\newcommand\copyrighttext{%
  \footnotesize \textcopyright \the\year{} IEEE. Personal use of this material is permitted. Permission from IEEE must be obtained for all other uses, including reprinting/republishing this material for advertising or promotional purposes, collecting new collected works for resale or redistribution to servers or lists, or reuse of any copyrighted component of this work in other works.}

\newcommand\copyrightnotice{%
\begin{tikzpicture}[remember picture,overlay]
\node[anchor=south,yshift=10pt] at (current page.south) {\fbox{\parbox{\dimexpr0.75\textwidth-\fboxsep-\fboxrule\relax}{\copyrighttext}}};
\end{tikzpicture}%
}

\title{\LARGE \bf
An Investigation of Denial of Service Attacks on Autonomous Driving Software and Hardware in Operation
}

\author{Tillmann Stübler$^{1}$, Andrea Amodei$^{2}$, Domenico Capriglione$^{2}$, Giuseppe Tomasso$^{2}$, \\ Nicolas Bonnotte$^{1}$, Shawan Mohammed$^{1}$  
\thanks{*Tillmann Stübler and Andrea Amodei contributed equally to this work and are both considered first authors.}
\thanks{$^{1}$Akkodis Germany Solutions GmbH, Resarch Division,
Flugfeld-Allee 12, D-71063 Sindelfingen
        {\tt\small tillmann.stuebler@akkodis.com}
        {\tt\small nicolas.bonnotte@akkodis.com}
        {\tt\small shawan.mohammed@akkodis.com}
        }%
\thanks{$^{2}$Department of Electrical and Information Engineering, University of Cassino and Southern Lazio, Via G. Di Biasio - 03043 Cassino (IT) 
        {\tt\small andrea.amodei@unicas.it}
{\tt\small capriglione@unicas.it}
{\tt\small tomasso@unicas.it}
}%
}

\begin{document}

\maketitle
\copyrightnotice

\thispagestyle{empty}
\pagestyle{empty}

\begin{abstract}
This research investigates the impact of Denial of Service (DoS) attacks, specifically Internet Control Message Protocol (ICMP) flood attacks, on Autonomous Driving (AD) systems, focusing on their control modules. Two experimental setups were created: the first involved an ICMP flood attack on a Raspberry Pi running an AD software stack, and the second examined the effects of single and double ICMP flood attacks on a Global Navigation Satellite System Real-Time Kinematic (GNSS-RTK) device for high-accuracy localization of an autonomous vehicle that is available on the market.
The results indicate a moderate impact of DoS attacks on the AD stack, where the increase in median computation time was marginal, suggesting a degree of resilience to these types of attacks. In contrast, the GNSS device demonstrated significant vulnerability: during DoS attacks, the sample rate dropped drastically to approximately 50\% and 5\% of the nominal rate for single and double attacker configurations, respectively. Additionally, the longest observed time increments were in the range of seconds during the attacks.
These results underscore the vulnerability of AD systems to DoS attacks and the critical need for robust cybersecurity measures. This work provides valuable insights into the design requirements of AD software stacks and highlights that external hardware and modules can be significant attack surfaces.
\end{abstract}

\section{INTRODUCTION}
As the dawn of Autonomous Driving (AD) technology reshapes the transportation landscape, integrating advanced software and hardware systems in vehicles promises a future of enhanced safety, efficiency, and connectivity \cite{ADsurvey, ADsurvey2, ADsurvey3}. Autonomous Vehicles (AVs), equipped with sophisticated sensors and AI-driven decision-making capabilities, are no longer a mere vision of the future but an evolving reality. This rapid advancement, however, brings with it an equally pressing challenge: ensuring the cybersecurity of these systems.

AVs rely heavily on interconnected systems for navigation, control, and communication, so they become potential targets for cyberthreats \cite{related_works14}, including a wide range of malicious
activities, each designed to exploit vulnerabilities in systems or networks, to steal or manipulate data, deny
a service or gain unauthorized access \cite{inf_sec_book2}. 
Denial of Service (DoS) and Distributed Denial of Service (DDoS) attacks represent significant risks. These attacks, characterized by their intent to overwhelm a system's resources and disrupt its normal functions, pose a severe threat to the reliability and safety of AVs.
In AD, the impact of DoS attacks is particularly concerning. Such attacks could impair vehicle communication systems, disrupt control modules, and compromise safety-critical functions that rely on the real-time capability of information. Despite this risk, little research explicitly addresses the resilience of AD systems to DoS attacks. This gap signifies a critical need for a focused investigation into AV cybersecurity's vulnerabilities and defense mechanisms.
This research aims to bridge this gap by exploring the susceptibility of AD systems to Internet Control Message Protocol (ICMP) Flood, a form of flooding DoS attack \cite{related_works16}. By simulating these attacks in controlled environments, this study seeks to understand the ramifications of such cybersecurity threats on the operational integrity of AVs. The objectives are twofold: first, to assess the impact of DoS attacks on different configurations of AD software and hardware; second, to derive insights that can inform the development of more robust cybersecurity measures for AVs.

\subsection{Real-Time Operation as a Hard Condition}
Automated driving systems necessitate real-time operation, where each process must be completed within a predefined time frame to ensure stability and safety. These systems, comprising various hardware and software components, face cumulative latencies that can critically delay responses to safety events. Achieving strict real-time performance is challenging due to the inherent complexities in modern software and hardware interactions. These complexities include high-level I/O abstractions integral to operating systems like Linux, compromising real-time performance due to CPU execution context switches during I/O operations. Furthermore, prevalent communication standards like TCP inherently impede strict real-time functioning. Consequently, a pragmatic approach often involves adopting ``soft" real-time bounds with adequate design margins to accommodate the unpredictable delays in the data processing chain.

\subsection{Autonomous Driving Software Stack}
The AD software stack is designed to determine control actions such as steering, accelerating, and braking, tailored to an automated vehicle's specific objectives. These actions are based on the vehicle's current state and its perception of the environment. At the heart of this stack is a controller, which incorporates various optimality criteria, including advanced functionalities like path planning and collision avoidance.

Our work uses a Model Predictive Control (MPC) \cite{MPC} approach. This MPC, closely integrated with higher-level functions, aims to broaden the concept of ``optimal control". It seeks to achieve a more comprehensive form of optimality by considering factors such as traffic conditions, the vehicle's surroundings, and the preferences and intentions of passengers. That way, several software modules commonly integrated in a loosely coupled way are being absorbed into a single Nonlinear Programming solver. However, the tightly-coupled nature of this component eliminates internal software and hardware boundaries which normally help to reduce the system's vulnerability to cyberattacks.

The solver finds the optimal solution of the combined objective functions and constraints imposed by the individual modules. Technically, at runtime, this MPC framework is a compiled software module that includes large amounts of generated code derived from symbolic expressions. This software was developed at Akkodis.

Although the MPC module satisfies real-time requirements, there are vulnerabilities because it needs to communicate with other software and hardware components to retrieve feedback from sensors and provide control signals to actuators. Except for these interfaces, the module does not perform any I/O operations nor allocate memory dynamically. We assume the principal vulnerabilities are I/O related to the mentioned communication interfaces and any non-exclusive access to the CPU when managed by the Linux kernel.

\subsection{Localization Module for AD}
Global Navigation Satellite System (GNSS) positioning is widely used to obtain accurate position and orientation information for automated vehicles. Real-Time Kinematic (RTK) positioning is an approach to obtain position estimates of predictable and uniform centimeter-level accuracy. This requires a reference GNSS receiver at a fixed and known location within a few kilometers of the rover (vehicle). Two essential concepts in RTK are the cancellation of atmospheric delays using correction data from the reference station and the utilization of highly accurate carrier phase measurements. The precondition for this is a reference station with a permanent data link to the rover. This, in turn, comes with vulnerability to cyberthreats.
We used a commercially available RTK GNSS system obtained from ANavS, a reference station, and survey antennas from the same supplier. The data link between the reference station and the rover was established via internal cellular mobile network modems. A Raspberry Pi is used internally to run the navigation software in both the rover and the reference station. We have little knowledge about software and kernel configuration in this device.

\subsection{DoS as a Potential Threat}

One of the most disruptive types of cyberthreats is represented by DoS and DDoS attacks. The first one is a malicious attempt to disrupt the expected behavior of a targeted server, user, or network by overwhelming it with a flood of illegitimate traffic \cite{inf_sec_book1}. On the contrary, a DDoS is an advanced form of DoS attack where multiple malicious systems, often referred to as botnets or zombies, are used to flood the target with traffic simultaneously. These attacks aim to make the targeted resource unavailable to its intended users. A holistic overview of DoS/DDoS cyberattacks may involve their categorization according to the TCP/IP protocol. Starting from the bottom, the Link Layer, MAC Flooding and ARP Spoofing are the more known. Regarding the Network Layer, there are ICMP Flood, and Smurf Attacks. After that, the SYN Flood and UDP Flood hit the Transport Layer. In the end, on the top of the stack, such as the Application Layer, the most known are represented by HTTP Flood and DNS Amplification attacks. 

\section{Related Work}
In recent years, integrating AD technology into modern vehicles has gained significant attention due to the increased risk of cyberattacks targeting critical components of autonomous vehicles and their infrastructure (V2X) \cite{related_works12}, which also ranges from control and planning to perception and localization. 
As reported in \cite{related_works1}, \cite{related_works11}, the surface of attacks is very unpredictable, considering that the malicious activity may come from different parts of the vehicle infrastructure in which every layer of the ISO model may be affected by a Denial of Service attack,\cite{related_works10}.
Indeed, due to the complexity of the environments, different scenarios are considered to evaluate the consequences of cyberattacks. In \cite{related_works2}, the authors proposed a network intrusion detection system to detect DDoS attacks in Vehicle-to-Vehicle (V2V) communication launched by several nodes to prevent the vehicle victim from getting legitimate resources.
However, possible malicious traffic may be generated by different vehicles in the Vehicle-to-Infrastructure (V2I) environment \cite{related_works3}.
As highlighted in \cite{related_works4}, a spamming DoS attack may introduce a non-predictable delay in the communication between vehicle and infrastructure, compromising the data availability in CIA triad \cite{related_works9}.   
Similarly \cite{related_works5}, the aim of a flooding attack, launched by the Roadside Unit (RSU) node, is to hit the bandwidth of the infrastructure, making it difficult or even impossible to reach the resources.

From this wide range of potential vulnerabilities, we focus specifically on real-time geolocation systems. 
This system is used with other sensor devices to enable precise localization of the AV, which is essential for autonomous driving. Therefore, adequate protection against potential cyberthreats must be ensured, \cite{related_works13}.
In particular, a DoS/DDoS attack can disrupt communications between the GNSS rover, the GNSS reference station (if used), and/or any other device communicating with the GNSS device. Loss of correction data will progressively degrade the rover's localization performance, which will continue to use outdated corrections \cite{related_works6}. Disruption of communications between the GNSS device and the vehicle control unit could immediately break closed-loop feedback control and render the AD system inoperable \cite{related_works8}.
In conclusion, integrating AD technology presents significant challenges in ensuring cybersecurity resilience against various cyberthreats, including DoS and DDoS attacks.

Moreover, the critical role of the real-time geolocation system in autonomous vehicles must be considered, as it is essential for accurate navigation and safe operation.

\section{Methodology}
\begin{figure}[!t]
    \centering
    \begin{subfigure}{0.4\textwidth}
        \centering
        \includegraphics[width=\linewidth]{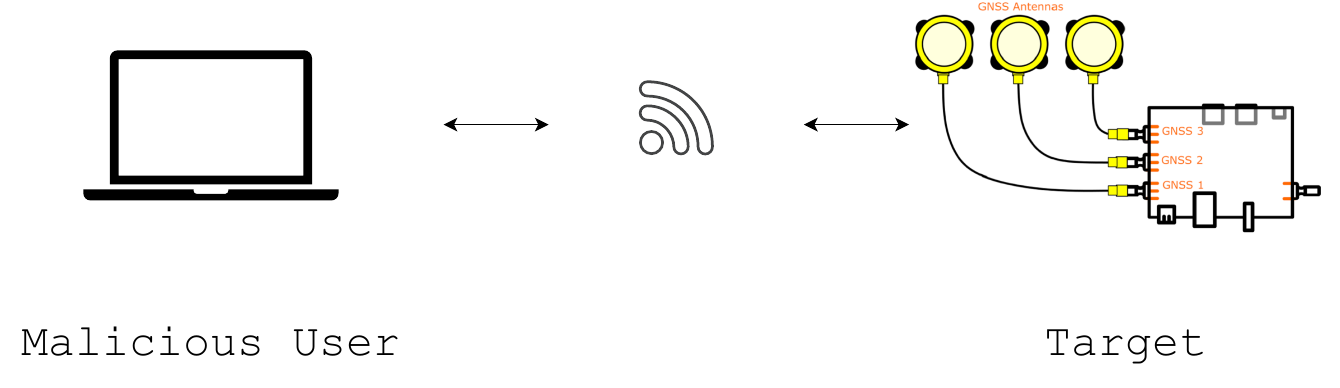}
        \caption{One-system attacker.}
        \label{fig:sub1_adas}
    \end{subfigure}
    \hfill
    \begin{subfigure}{0.4\textwidth}
        \centering
        \includegraphics[width=\linewidth]{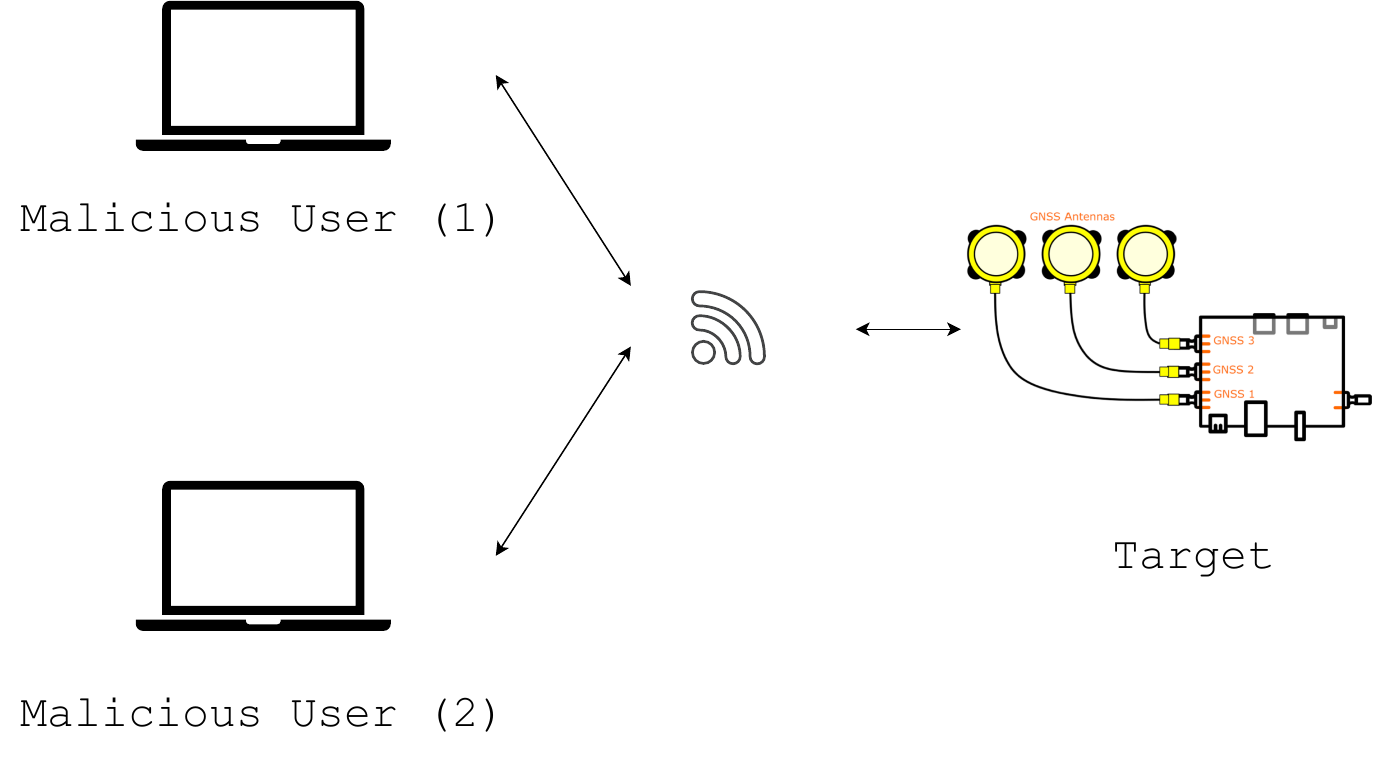}
        \caption{Two-system attackers.}
        \label{fig:sub2_adas}
    \end{subfigure}
    \caption{Overview of physical setup involving the attacker, performing an ICMP Flood attack, connected via wifi to a GNSS-RTK device.}
    \label{fig:setup_adas}
\end{figure}

In this work, we perform an ICMP Flood attack to test the performance of the physical setup; in detail, this challenge aimed to saturate network bandwidth and exhaust CPU or I/O resources. In a typical system, ICMP is commonly used for diagnostic and control purposes within IP networks, including error reporting and network management. 
A standard ICMP header packet comprises five fields: Type, code, checksum, rest of header, and payload. The first three are made up of 8 bits, the checksum field has 16 bits, the fifth may have up to 32 bits, and the payload has a variable length.
It is possible to overwhelm a device by sending a large volume of these packets without waiting for the reply, leading to congestion and slowing down or disrupting legitimate network traffic. Various physical setups are used to carry out this attack, shown in Figures \ref{fig:setup1_adas} and \ref{fig:setup_adas}. 
\subsection{DoS Attack on AD Stack}
\begin{figure}[t]
    \centering
    \includegraphics[width=0.4\textwidth]{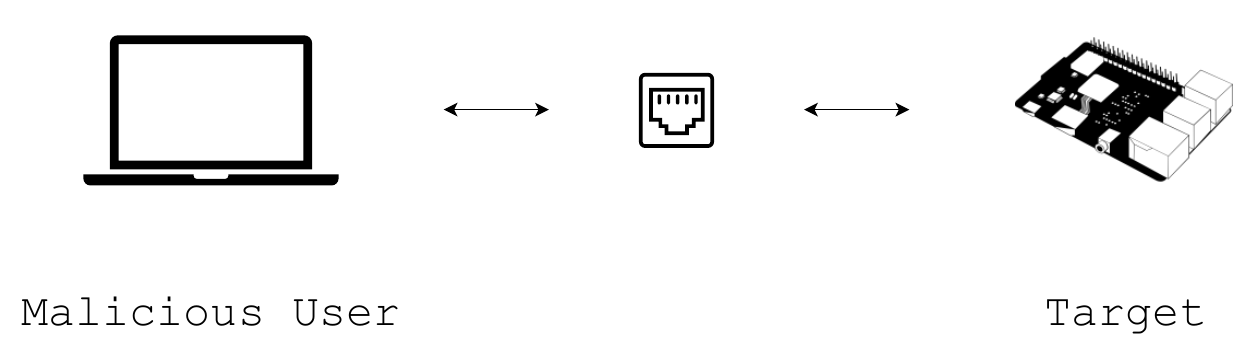}
    \caption{Overview of physical setup involving the attacker, performing an ICMP Flood attack, connected via ethernet to a Raspberry Pi running an MPC motion controller module.}
    \label{fig:setup1_adas}
\end{figure}

First (Fig. \ref{fig:setup1_adas}), an ICMP flood attack is carried out from a MacBook Pro mounted by an Apple M2 Max Chip connected via ethernet Cat.6 cable supporting a data rate of 1Gbps. The ICMP Flood is launched from a virtual environment in Kali Linux where each header's packet has 28 bytes (8 bytes indicates the header size of the ICMP packet and 20 bytes for the IP header).  

On the other side, a Raspberry Pi 4 Model B, with a Gigabit Ethernet interface, running the AD software stack was the target.
The Raspberry has a standard installation of Raspberry Pi OS, and various communication, logging, and monitoring functions are implemented in Python.  Communication with external components (feedback and control signals) is replaced by a low-footprint vehicle simulator implemented in Python. The MPC framework is compiled as a single Python extension module.
\begin{table}[htb]
\begin{tabularx}{\linewidth}{|l|X|}
\hline
Host hardware & Raspberry Pi 4B \\
Host OS & Raspberry Pi OS (Bookworm) \\
ADAS Software & proprietary ADAS MPC stack, malloc-free, compiled with Clang as Python extension module \\
Network & ethernet cable 6Cat. (no switch / router involved) \\
Size ICMP Packets & 28 header bytes \\
Rate & approx. 300K packets per second \\
Test procedure & 50,000 iterations of MPC motion controller \\
Data acquisition & computational latency measured using clock\_gettime(CLOCK\_MONOTONIC, \textellipsis) \\
\hline
\end{tabularx}
\caption{AD stack bench setup}
\label{table_adas}
\end{table}

During the test, system time when entering and exiting the compiled extension module of the MPC is recorded via the function \textit{clock\_gettime(CLOCK\_MONOTONIC, \textellipsis)} of the C standard library, which wraps the corresponding Linux system call. On the Raspberry Pi, the kernel implementation defers to a tick counter register of the BCM2711 with the counter clocked at a fixed 1 MHz rate. Thus, we measure durations in actual time spent at a resolution of 1µs. All the results are logged into files. The measurements do not include time spent in the Python interpreter (which was used for convenience only). If the process is preempted by the operating system while calculating the MPC update, this delay is included in the measurement. 50,000 iterations were carried out with the Raspberry Pi under DoS attack and without an attack for reference. It should be noted that closing the feedback loop with a simulated vehicle avoids the system calls typically associated with hardware I/O operations, so we benchmarked the impact of the DoS attack on software real-time performance only. However, every \textit{clock\_gettime} call incurs one system call with possible delays.

\subsection{DoS Attack on GNSS-RTK device}
In the second scenario, the authors benchmarked a GNSS device of ANavS GmbH under a DoS attack. Unlike the previous scenario, a wifi connection is used for the attack, leveraging on the access point delivered by the ANavS module. We tested the effects of one (Fig. \ref{fig:sub1_adas}) and two malicious users (Fig. \ref{fig:sub2_adas}).
Tab. \ref{table_gnss} summarizes this scenario. The DoS attack is launched from either one (configuration 1) or two (configuration 2) laptops. In both cases, the ICMP flood attack on the target device is initiated at $t = 10 s$. In each configuration, the test is repeated 10 times.
\begin{table}[htb]
\centering
\begin{tabularx}{\linewidth}{|l|X|}
\hline
Host hardware & Raspberry Pi (unknown model) \\
Host OS & Raspberry Pi OS (unknown release) \\
Network & Wifi - access point of GNSS module \\
Size ICMP Packets & 28 header bytes \\
Rate (configuration 1) & approx. 500K packets per second \\
Rate (configuration 2) & approx. 1000K packets per second \\
Test procedure & 10 repetitions of 30s tests, with DoS attack at $t = 10 \ldots 30 s$ \\
Data acquisition & Streamed data (device-specific protocol) recorded to file, GPS and system timestamps extracted for analysis\\
\hline
\end{tabularx}
\caption{GNSS device bench setup}
\label{table_gnss}
\end{table}

\section{Results}

\subsection{Attack on AD Software Stack}
\begin{figure}[tbh]
    \centering
    \includegraphics{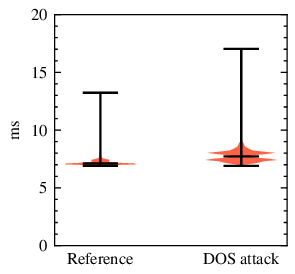}
    \caption{AD stack computation time per time step, with horizontal bars at minimum, median, and maximum value}\label{fig:adas_stats}
\end{figure}

Statistics of recorded timing measurements were calculated to assess the attack impact on the AD stack's real-time performance. The median time spent in the MPC update function was approx. 7 ms in the reference (reference), and only slightly increased during the DoS attack. Standard deviation was well below 1 ms in both cases. Infrequently, additional delays were observed, most likely due to the process being preempted. The longest durations were approx. 13 ms in the reference vs. 17 ms with the system under DoS attack. Figure \ref{fig:adas_stats} shows violin plots of computation time for both scenarios compared. The empirical distributions are narrow. The lower 99 percentiles lie within a 1.2 ms (reference) and 2.6 ms (DoS attack) interval, respectively.

\subsection{Attack on GNSS Device}
Our impact analysis is based on binary data streamed on port 6001 of the GNSS device, which we dumped to files on a PC during all tests. The duration of every test was 30s. For both configurations (1 DoS attacker, or 2 simultaneous attackers), we carried out 10 repetitions and captured all data. The data stream was later decoded using a custom implementation of the device-specific binary protocol. The binary packages include the position and orientation solutions, along with various diagnostic data.

At the beginning of every recording, the GNSS device operated normally. We initiated the DoS attacks at $t = 10 s$, potentially resulting in performance degradation. For every recording, we calculated characteristic values separately for a snippet of data from 0s until 8s (reference snippet with no DoS attack) and for a snippet from 12s until 30s (with DoS attack). We intentionally discarded the data between 8s and 12s because the timed initiation of recordings and DoS attacks is affected by minor uncertainties.

\begin{figure}[tbh]
    \centering
    \includegraphics{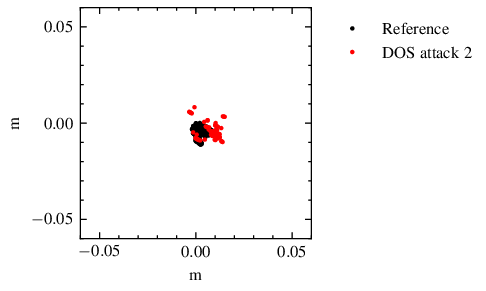}
    \caption{Positions in horizontal plane during one 30s experiment, while antennas and receiver were not moving}\label{fig:position}
\end{figure}

First, we screened the position solutions in all recordings for plausibility. The position solution was found plausible and converged in all recordings in the subsequent analysis. Figure \ref{fig:position} shows, as an example, the position solutions of one recording of configuration 1. Visualization is in a local east/north coordinate system with the origin at the first position solution recorded.

Then, for both configurations, we merged the 10 recordings and calculated statistical values on the combined datasets. In order to analyze the impact of the DoS attacks on the timing of the position solution and the data packets, our analysis relies on two data fields in the binary packets:
\begin{itemize}
\item the sampling timestamp of the solution, expressed according to GPS ``Time of Week" convention (the field is termed ``tow")
\item the timestamp of when the particular solution was processed by the GNSS device, according to its internal clock (this field is called ``systemTimeSolOut")
\end{itemize}

\begin{figure}[tbh]
    \centering
    \includegraphics{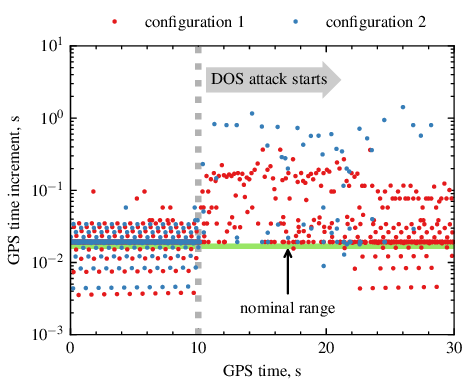}
    \caption{Sample-to-sample time increments in one recording (configuration 2)}\label{fig:timing}
\end{figure}

In our tests, we configured the GNSS device with a ``medium output rate", corresponding to a sample rate between 55 Hz and 65 Hz. Figure \ref{fig:timing} shows sample-to-sample time intervals obtained from one single recording of configurations 1 and 2. This GNSS device clearly operates in a nonuniform sampling mode for $t < 10 s$ (no DoS attack). This uncommon behavior is the result of a data fusion algorithm processing two incoherently sampled data streams - inertial measurements and GNSS observables from different hardware devices. Furthermore, the fusion algorithm occasionally discards an IMU update in favor of an upcoming GNSS update. This results in individual time increments spanning almost the interval from zero to twice the nominal time increment. Occasionally, increments are way beyond this upper bound, indicating this device does not satisfy real-time requirements.

\begin{figure}[tbh]
    \centering
    \includegraphics{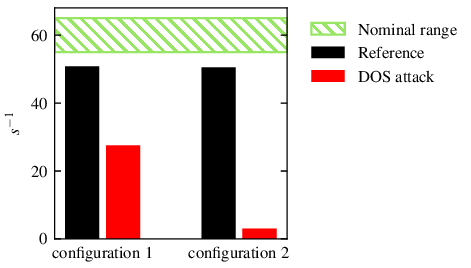}
    \caption{Mean localization sample rate}\label{fig:rate_bars}
\end{figure}

For $t > 10 s$, sampling is erratic in both configurations but more pronounced in configuration 2. Figure \ref{fig:rate_bars} shows the mean sample rates (referring to sampling timestamps) for the two configurations, with a comparison of the attack and reference phases. For the reference phase, the observed mean rate was slightly below the nominal range. During the DoS attack, the sample rate dropped to approximately 50\% and 5\% of the lower end of the nominal range, respectively.

\begin{figure}[tbh]
    \centering
    \includegraphics{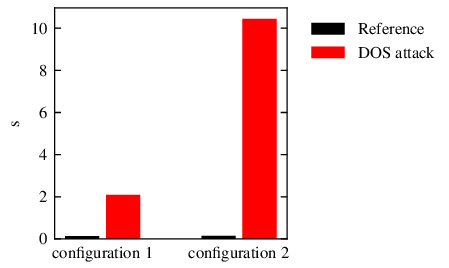}
    \caption{Longest GNSS localization outages across the reference and DOS recordings}\label{fig:outage_bars}
\end{figure}

Figure \ref{fig:outage_bars} shows the longest observed time increments in the respective dataset, which are in the range of seconds for both DoS attack configurations. Again, the figures refer to sampling time, so any time increment exceeding the nominal range indicates samples dropped (rather than delayed).

\begin{figure}[tbh]
    \centering
    \includegraphics{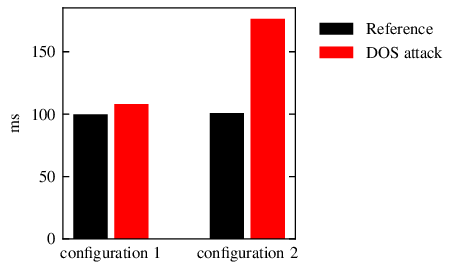}
    \caption{Double difference jitter (Eq. \ref{eq:dd_jitter}) in reference and DOS recordings}\label{fig:jitter_bars}
\end{figure}

Finally, we estimated the the double difference jitter $T_{dd}$ according to
\begin{equation}
\begin{split}
t_{dd,i} = t_{tow,i} - t_{tow,i+1} + t_{sys,i+1} - t_{sys,i} \\
T_{dd}(\boldsymbol{t}_{tow}, \boldsymbol{t}_{sys}) = Q_{0.95}(\boldsymbol{t}_{dd}) - Q_{0.05}(\boldsymbol{t}_{dd})
\end{split}
\label{eq:dd_jitter}
\end{equation}
where $Q_p$ denotes the $p$-Quantile, $\boldsymbol{t}_{tow}$ and $\boldsymbol{t}_{sys}$ the series of tow and systemTimeSolOut values unpacked from the recorded binary packages. The GNSS tracking loop guarantees that tow (sampling time) always corresponds to the actual time of arrival of a particular snippet of RF waves; It is unaffected by the CPU solving the navigation equations. The field systemTimeSolOut is filled with the current system time (which is not GPS time) just before transmitting the latest position solution. In the double differences $T_{dd}$, both the constant offset of GPS vs. system time and any constant processing latency are being canceled. Thus, $T_{dd}$ is nonzero only if processing latency changes from one sample to the next. The results (Fig. \ref{fig:jitter_bars}) show a considerable variation (approx. 0.1s) even in the reference data. However, this variation is not significantly elevated during the DoS attacks. There is no discrepancy in sample rates and jitter estimates in the reference phases of configuration 1 vs. 2.

\section{DISCUSSION AND CONCLUSIONS}
In our study, we comprehensively evaluated a ``soft" real-time system's resilience to DoS attacks, focusing on two scenarios involving a Raspberry Pi platform. It is important to note that while Raspberry Pis are not typically associated with high-reliability real-time applications, their use in this context offers valuable insights into the performance of less robust systems under cyberattack conditions.
\paragraph{Resilience of AD Software Stack} Our findings in the first scenario were unexpected. Despite running on a non-hardened Linux installation without real-time kernel enhancements or specific OS optimizations, the AD software stack exhibited a surprising degree of robustness compared to the presumably more fortified GNSS-RTK device. The AD stack maintained acceptable worst-case delays throughout 50,000-time steps, which roughly translates to 900 seconds in standard application scenarios, even under DoS attack conditions. This is particularly notable considering the system's intended update rate of 55 Hz, underscoring its potential for reliable performance even in less-than-ideal operational environments.
\paragraph{Vulnerability of GNSS-RTK Device}
Conversely, the GNSS-RTK device was significantly impacted during the DoS attacks, effectively ceasing normal operations. This differential impact could be attributed to the GNSS navigation algorithm's reliance on real-time communications with various GNSS hardware receivers, a factor absent in the AD stack. The latter's relative insulation from I/O congestion effects, due to its closed feedback loop through a simulation model, may also contribute to its resilience.
\paragraph{Implications of Jitter Estimates}
Interestingly, our analysis of jitter estimates (refer to Figure \ref{fig:jitter_bars}) revealed that the GNSS-RTK device's operation during the DoS attacks resulted in more data loss than delayed processing. This behavior aligns with operational logic in closed-loop control applications, where outdated system state information is less valuable than more current data. Consequently, the system prioritizes processing the latest data over catching up on delayed information.
\paragraph{Insights into System Bottlenecks}
The recorded data suggests that the primary challenges faced by the system under DoS attack conditions are related more to I/O bottlenecks rather than CPU limitations. This inference is supported by the observed jitter patterns, which indicate that the navigation software's performance degradation is not primarily due to process preemption. An increased preemption rate would likely result in a significant rise in double difference jitter, as described by Equation \ref{eq:dd_jitter}. This suggests that the main issue is likely the kernel's inability to effectively prioritize a high volume of individual I/O events under stress.
\paragraph{Real-world Applicability and Cybersecurity Considerations}
Finally, the practicality of such DoS attacks in real-world scenarios warrants consideration. Our experimental setup, utilizing either wired or wireless local networks, allowed for significant bandwidth usage. However, replicating this attack scale via the internet would necessitate a large coordinated network of malicious machines. Additionally, cybersecurity measures such as firewalls can mitigate these risks. Nevertheless, the GNSS device, which requires a constant data connection for correction data, is a persistent vulnerability even if robust protection measures are in place. The AD Stack, on the other hand, may require vehicle-to-vehicle or vehicle-to-infrastructure communications via the internet.

In summary, our study highlights the varying levels of resilience in soft real-time systems under DoS attack conditions. The AD software stack's performance, despite its non-optimized setup, contrasts with the GNSS-RTK device's vulnerability, offering insights into system design and cybersecurity strategies for real-time applications.

\addtolength{\textheight}{-12cm}   




\bibliographystyle{IEEEtran}

\bibliography{biblio}

\end{document}